\documentstyle[preprint,aps]{revtex}

\newcommand{\be}{\begin{equation}}
\newcommand{\ee}{\end{equation}}
\newcommand{\bear}{\begin{eqnarray}}
\newcommand{\ear}{\end{eqnarray}}
\newcommand{\ba}{\begin{eqnarray*}}
\newcommand{\ea}{\end{eqnarray*}}
\renewcommand{\theequation}{\arabic{section}.\arabic{equation}}

\begin{document}
\begin{titlepage}
\setcounter{page}{1}
\begin{flushright}
HD--THEP--98--15\\
\end{flushright}
\vskip1.5cm
\begin{center}
{\large{\bf Hilbert Space of Isomorphic Representations of}}\\
{\large{\bf Bosonized Chiral $QCD_2$}}\\
\vspace{1cm}
{L. V. Belvedere\footnote{email: belve@if.uff.br}, R. L. P. G. do Amaral }\\
{\it Instituto de F\'{\i}sica - Universidade Federal Fluminense}
\\
{\it Av. Litor\^anea, S/N, Boa Viagem, Niter\'oi,
CEP.24210-340, Rio de Janeiro - Brasil}
\\
K. D. Rothe \footnote{email: k.rothe@thphys-uni-heidelberg.de}
\\
{\it Institut  f\"ur Theoretische Physik - Universit\"at Heidelberg}
\\
{\it Philosophenweg 16, D-69120 Heidelberg}
\\
{(April 1, 1998)}
\end{center}

\begin{abstract}
{\small We analyse the Hilbert space structure of the isomorphic
gauge non-invariant and gauge invariant bosonized formulations of chiral
$QCD_2$ for the particular case of the Jackiw-Rajaraman
parameter $ a = 2$. The BRST subsidiary conditions are found not to provide a
sufficient criterium for defining physical states in the Hilbert
space and additional superselection rules
must to be taken into account. We examine the effect of the use of a
redundant field algebra in deriving basic
properties of the model. We also discuss the constraint structure of
the gauge invariant formulation and show that the only primary
constraints are of first class.}
\end{abstract}
\end{titlepage}
\newpage
\section{Introduction}

In a series of papers \cite{AA1,CRS,CR,AR,RST} the Hilbert space and structural
properties of bosonized $QCD_2$ in the so-called \cite{AA1} ``local''
and ``non-local'' decoupled formulation have been extensively discussed. In particular
it has been shown \cite{AR} that the BRST conditions associated with the
change from fermionic to bosonic variables play the role of the familiar
Lowenstein-Swieca conditions \cite{LS} defining the physical Hilbert space
in the abelian case. This analysis has recently been extended to chiral
$QCD_2$ \cite{ABRS} in the so-called ``gauge non-invariant'' (GNI) formulation, where it was shown that the Becchi-Rouet-Stora-Tyutin (BRST)
conditions play the role of the
conditions of Boyanovski et al \cite{B} in the abelian case. In
ref. \cite{ABRS} it was further argued, relying on previous considerations
in the abelian case \cite{Bel}, that chiral $QCD_2$ for the
Jackiw-Rajaraman (JR) parameter $a=2$ \cite{JR,GRR} is in fact not equivalent
to $QCD_2$, contrary to what superficially appears to be the case \cite{CW}.

In the present paper
we complete this demonstration, by examining in detail the physical
Hilbert space ${\cal H}_{phys}$ of the non-Abelian bosonized formulation,
as defined by a set of BRST conditions.
As pointed out in Refs.\cite{Bel,CBAL,MPS,S}, the use of bosonization
techniques  raises some
delicate questions related to the use of a redundant Bose field
algebra. This field algebra contains more degrees of freedom than
those needed for the
description of the model. Since for an anomalous theory the gauge invariance is spoiled at the quantum level, the physical content of the
theory relies strongly and directly on the field algebra intrinsic to the
model \cite{Bel}. Taking proper care in the construction of
the Hilbert space associated with the Wightman functions that define
the theory, and examining the ``charge'' content of the Hilbert space,
we find that the BRST conditions
are not sufficient to define ${\cal H}_{phys}$, and that certain superselection
rules have also to be respected. We theirby show that the Hilbert
space of chiral $QCD_2$ defined for $a = 2$ does not
contain the Hilbert space of $QCD_2$ as a physical subspace. 
This is the content of section 2.
In section 3 we then turn to the so-called  ``gauge invariant'' (GI) 
formulation
\cite{GR} obtained by embedding the GNI bosonized formulation into
a gauge theory, following the standard procedure of refs. \cite{HaTsu}.
We show in particular that in this case there exist in addition to the
above BRST constraints, precisely one set of (primary) constraints
in the sense of Dirac \cite{Di}. We show these constraints to be first class, and to be the generators of the gauge symmetries introduced by the
gauge invariant embedding, as expected. Some further considerations and
technical details relating to sections 2 and 3 are relegated to the
appendices A and B, respectively.

\section{Gauge Non-Invariant Local Formulation}

In Minkowski space, the generating functional of the $GNI$ formulation of
chiral $QCD_2$ with
left-moving fermions coupled to a $SU(N)$ gauge field is given by
$$
Z\, \big [ \overline \eta,\eta, J_\mu \big ]\,
=\, \int {\cal D} {\cal A}_\mu\,\int {\cal D} \psi^o_r\,
{\cal D}\psi_r^{o\,\dagger}\,\int {\cal D} \psi_\ell\,{\cal D}\psi_\ell^\dagger\,
\exp \Big \{\,i\,S[{\cal A}_\mu,\psi,\overline\psi] \Big \}\,\times
$$
\be
\times\,\exp \Big \{\, i \int\,d^2 z\,\Big (\,
\eta^\dagger_{_r} \,\psi_r^o\, +\, \psi_r^{o\,\dagger}\,\eta_{_r}\, + \,
\eta^\dagger_{_\ell}\,\psi_\ell \,+\, \psi_\ell^\dagger\, \eta_{_\ell}\, +\, J_\mu\,
 {\cal A}^\mu\,\Big ) \Big \}\,,
\ee
with\footnote{\it Our conventions are: $A_\pm = A_0\pm A_1, \partial_{\pm} =
\partial_0\pm\partial_1, \epsilon^{01} = 1$.}
\be S[{\cal A}_\mu,\psi,\overline\psi]\,=\,\int d^2\,z\,
\left\{ - \frac{1}{4}\,{tr}\,F^{\mu\nu} F_{\mu\nu}\,+\,
\psi_r^{o\,\dagger}i\partial_+\,\psi_r^o\,+\,\psi^\dagger_\ell\,
(i\partial_- + eA_- )\,\psi_\ell\, \right\}\,.
\ee

Parametrizing $ {\cal A}_\pm $ in terms of $SU(N)$ matrix valued fields as
follows\footnote{\it We follow
here the notation of Refs.\cite{CRS,CR,ABRS}.}
\be
e{\cal A}_+ = U^{-1} i \partial_+ U\,\,,\,\,e{\cal A}_- = V i \partial_- V^{-1}\,,
\ee
making the change of
variables $ {\cal A}_+ \rightarrow U $, $ {\cal A}_- \rightarrow V $, as
well as the chiral rotation
\be
\psi_\ell^o = V^{-1} \psi_\ell\,,
\ee
and taking due account of the Jacobians in the integration
measure \cite{FNS,AA1}, we arrive, following the procedure of
references \cite{AA1,CRS}, at the generating functional \cite{ABRS}
\be\label{hatZ}
 Z\,\big [\overline \eta,\eta,J_\mu \big ] =  {\cal Z}_{gh}^o\,\,
\hat{\cal Z}\,\big [\overline \eta,\eta,J_\mu \big ]\,,
\ee
where $ {\cal Z}_{gh}^{(0)} $ is the partition function of free ghosts \cite{CRS}
associated with the change of variables (2.3),
\be\label{Zgh}
Z_{gh}^{(0)}=\int{\cal D}b^{(0)}_+{\cal D}c_+
^{(0)}e^{i\int tr b_+^{(0)}i\partial_-c_+^{(0)}}
\int{\cal D}b_-^{(0)}{\cal D}c_-^{(0)}e^{i\int tr b_-^{(0)}i
\partial_+c_-^{(0)}},
\ee
$\hat {\cal Z}$ is given by

\bear\label{generatingZ}
 \hat{\cal Z}\,\big [\overline \eta,\eta,J_\mu \big ] &&= \int{\cal D} \psi^o \,
{\cal D} \overline\psi^o\, e^{\,i\,\int\,d^2 z\,\Big ( \overline\psi^o\,i\,
\slash \!\!\!\partial\,\psi^o + \eta^\dagger_{_r} \psi^o_r +
\psi_r^{o\,\dagger} \eta_{_r} \Big ) }\nonumber\\
&&\times \int\,{\cal D} U\,{\cal D} V\,e^{\,iS [UV] +
i \int\,d^2 z\,\Big \{\,\eta^\dagger_{_\ell}\,V\,\psi^o_\ell \,+\, \psi_\ell^{o\,
\dagger}\,V^{-1}\,\eta_{_\ell}\, +\,\frac{1}{e} J_-\,
[ U^{-1}\, i\,\partial_+ U ] \,+\,\frac{1}{e}\,J_+\,
 [ V \,i\,\partial_- V^{-1} ]\,\Big \} }\,,
\ear
and the effective bosonized action is
\be\label{efbosaction}
S [U,V] = S_{_{_{YM}}}[\Sigma]\,-\,\Big ( C_{_V} + \frac{a}{2} \Big)\,
\Gamma[U V] \, + \frac{a}{2} \Gamma [U] + \Big (\frac{a}{2} - 1
\Big ) \Gamma [V]\,.
\ee
Here
\bear\label{2.4}
S_{YM}[\Sigma]&=&\frac{1}{4e^2}\int {tr}\frac{1}{2}
[\partial_+(\Sigma i\partial_-\Sigma^{-1})]^2\\
&=&\frac{1}{4e^2}\int {tr}\frac{1}{2}
[\partial_-(\Sigma^{-1}i\partial_+\Sigma)]^2.
\ear
is the Yang-Mills action with $\Sigma$ the
gauge-invariant variable $\Sigma=UV$, and
$ \Gamma [G] $ is the Wess-Zumino-Witten action \cite{WZW}.
Changing variables from $ \{U,V\} $ to $ \{\Sigma,U\} $, we have \cite{ABRS}
\be\label{GNIA}
S [U,V] \rightarrow S [U,\Sigma] = S_{_{_{YM}}}[\Sigma ]\,-\,
\Big ( C_{_V} + \frac{a}{2} \Big )\,\Gamma[\Sigma ] \,+\,
 \frac{a}{2} \Gamma [U] + \Big (\frac{a}{2} - 1 \Big ) \Gamma [U^{-1}
\Sigma]\,.
\ee
This action provides the starting point for the discussion to follow.

\subsection{FIELD ALGEBRA AND HILBERT SPACE}

The equations of motion defining the $GNI$ formulation of chiral
$QCD_2$ are given in terms of the fundamental set of field
operators $\{\psi^o_r,\psi_\ell,{\cal A}_\mu\}$. Within the context
of general principles of Wightman field theory, these field
operators constitute the intrinsic mathematical description of the
theory and serve as the building material in terms of which the $GNI$
version of the model is formulated \cite{Bel}. The set of field operators
$\{\psi^o_r,\psi_\ell,{\cal A}_\mu\}$ defines a local field algebra
$\mbox{\boldmath $\Im$}$, and the Wightman functions generated from
the field algebra $\mbox{\boldmath $\Im$}$ identifies a Hilbert space
$\mbox{\boldmath ${\cal H}$} \doteq \mbox{\boldmath $\Im$}
\,\mbox{\boldmath{$\Psi_o$}} $ of the $GNI$ formulation of the model.

The bosonization of the model requires the use of a
larger field algebra, which includes non-observable Bose fields as
well as ghosts. In the local formulation the resulting effective theory is
given in terms of the set of
fields $\{\psi^o_r,\psi^o_\ell,U,V,gh\}$. These field operators
generate an extended Bose-Fermi-ghost field
algebra $\mbox{\boldmath $\Im$}^{^e}$ which is represented in the
Hilbert space $\mbox{\boldmath ${\cal H}$}^{^e}$. The field algebra
$\mbox{\boldmath $\Im$}$ is a proper subalgebra
of $\mbox{\boldmath $\Im$}^{^e}$ and the Hilbert
space  $\mbox{\boldmath ${\cal H}$}$ is a
subspace of $\mbox{\boldmath ${\cal H}$}^{^e}$.

The field
algebra $\mbox{\boldmath $\Im$}^{^e}$ contains elements not
intrinsic to the model, which should not be considered as elements of
the intrinsic field algebra $\mbox{\boldmath $\Im$}$: not all fields
$ \{\psi^o_r,\psi^o_\ell,U,V,gh\} $ belong
to the algebra $\mbox{\boldmath $\Im$}$, nor all vectors
of $\mbox{\boldmath ${\cal H}$}^{^e}$ belong to the state
space  $\mbox{\boldmath ${\cal H}$} $. Nevertheless, combinations like $ V^{-1}
\psi^o_\ell$ and $U^{-1} \partial_+ U$, $V
\partial_-V^{-1}$, do evidently belong to the field algebra $\mbox{\boldmath $\Im$}$.

The physical Hilbert space $\mbox{\boldmath ${\cal H}$}_{_{phys.}}$ is a
representation of the field
algebra $\mbox{\boldmath $\Im$}_{_{phys.}}$ which satisfies the
subsidiary BRST condition
\be
\big [ {\cal Q}_{_{BRST}} , \mbox{\boldmath $\Im$}_{_{phys.}} \big ] = 0\,,
\ee
to be specified later on. Hence
\be\label{BRST}
{\cal Q}_{_{BRST}} \,\mbox{\boldmath ${\cal H}$}_{_{phys.}} = {\cal
Q}_{_{BRST}}\,\mbox{\boldmath $\Im$}_{_{phys.}}\,
\mbox{\boldmath{$\Psi_o$}} = 0\,.
\ee
These conditions correspond in the abelian case (chiral
Schwinger model) \cite{B,AR} to the familiar Lowenstein-Swieca \cite{LS}
conditions requiring that the longitudinal part of the current
annhilate the physical states.

As stressed in Refs.\cite{Bel,ABRS}, since the theory has lost the
local
gauge invariance at the quantum level, it is a peculiarity of the
anomalous chiral theory that $\mbox{\boldmath $\Im$}_{_{phys.}} \equiv
\mbox{\boldmath $\Im$} $; all
operators belonging to the intrinsic local field
algebra $\mbox{\boldmath $\Im$}$ commute with the BRST charges
 and thus represent
physical observables \cite{ABRS}. Hence $\mbox{\boldmath ${\cal H}$}
= \mbox{\boldmath ${\cal H}$}_{_{phys.}}$. The Hilbert space is thus
isomorphic to $\mbox{\boldmath ${\cal H}$} = \mbox{\boldmath ${\cal H}$}_{\psi^o_r} \otimes
\mbox{\boldmath ${\cal H}$}_{\psi_\ell} \otimes
\mbox{\boldmath ${\cal H}$}_{{\cal A}_\mu} $, contrary to what happens in a
genuine gauge theory like $QED$ and
$QCD$, in which the physical Hilbert subspace is defined by equivalence classes
corresponding to gauge invariant states. In terms of the bosonic field
variables $\{U,V\}$, or $\{\Sigma,U\}$, we can
write $ \mbox{\boldmath ${\cal H}$} \equiv
\mbox{\boldmath ${\cal H}$}_{\psi^o_r} \otimes
\mbox{\boldmath ${\cal H}$}_{_{\Sigma,U,\psi^o_\ell,gh}} $. The BRST
conditions (\ref{BRST}) impose non-trivial restrictions
on further decompositions of the
closure of the space $\mbox{\boldmath  ${\cal H}$}_{_{\Sigma,U,\psi^o_\ell,gh}}$.

The field algebra can be further enlarged by introducing the bosonized
partition function of free ($U(N)$) Fermi fields $\psi^o$ in the
factorized form
\be \label{ZFbosonized}
\int{\cal D} \psi^o \,{\cal D}
\overline{\psi}^o\, e^{\,i\,\int d^2z\, \overline{\psi}^o\,i\,
\slash \!\!\!\partial\,\psi^o}\, = \,{\cal Z}^{^{(0)}}_{_{U(1)}} \times\,
\int \,{\cal D} G\,e^{\,i \Gamma [G]}\,,
\ee
where, like $U$ and $V$, $G$ is a $SU(N)$ matrix valued field. The
factorized $U(1)$ partition function is given by
\be
{\cal Z}^{^{(0)}}_{_{U(1)}} = \int\,{\cal D}\,\Phi\,e^{\,\frac{i}{2}\,\int\,d^2 z\,
(\partial_\mu \Phi )^2}\,,
\ee
where the massless scalar field $\Phi$ acts as potential for the
conserved $U(1)$ current
\be
{\cal J}^\mu \,=\, - \,\frac{1}{\sqrt \pi}\, \partial ^\mu\,\Phi\,.
\ee
In this way, we obtain a field algebra $\mbox{\boldmath
$\Im$}^{^E}\,\supset\,\mbox{\boldmath $\Im$}^{^e} $, generated
from the set of field operators $\{U,V,G,\Phi,gh\}$, and represented in the enlarged Hilbert
space $\mbox{\boldmath ${\cal H}$}^{^E}\,\supset\,
\mbox{\boldmath ${\cal H}$}^{^e}$.

\subsection{THE CASE $a = 2$.}

For $ a = 2 $, the effective bosonized action (\ref{GNIA}) decouples
as \cite{ABRS}
\be
S [U,\Sigma] = S_{_{_{QCD_2}}}[\Sigma ]\,+\,\Gamma [U] \,.
\ee
Except for the ``decoupled'' WZW action $ \Gamma[U] $, which appears
to play merely a spectator role, this is the action of $QCD_2$ in the
decoupled local formulation \cite{AA1,CRS,CR,ABRS,RST}.The partition function thus factorizes as follows:
\be\label{factorizedZhat}
 Z\,\big [0\big ] = {\cal Z}^{^U}_{_{WZW}}\,\Big ( {\cal Z}^{^{(0)}}_{gh}
{\cal Z}^{^{(0)}}_{U(1)}{\cal Z}^{G}_{WZW} {\cal Z}^{^\Sigma }\Big )_{QCD_2}\,.
\ee
However, the factorization does not apply to the Hilbert space since the fields
$U$, $\Sigma$ and $\psi^o_{\ell}$ are coupled by the BRST conditions
(\ref{BRST})\cite{ABRS} with
\be\label{2.39}
{\cal Q}_{_{BRST}}={\cal Q}_\pm=\int dx^1{tr} c_\pm^{(0)} \Big (\Omega_\pm
-\frac{1}{2}\{b_\pm^{(0)},c_\pm^{(0)}\}\Big )\,,
\ee
where
\begin{eqnarray}\label{constr+}
\Omega_+ &=& \frac{1}{4e^2}\Sigma^{-1}[\partial_+^2
(\Sigma i\partial_-\Sigma^{-1})]\Sigma \nonumber\\
&-&\frac{1+C_V}{4\pi}
\Sigma^{-1}i\partial_+\Sigma
+\psi_l^{o}\psi_l^{o\dagger}+\{b^{(0)}_+,c_+^{(0)}\}
\end{eqnarray}
\begin{eqnarray}\label{constr-}
\Omega_- &=& \frac{1}{4e^2}\Sigma[\partial_-^2(\Sigma^{-1}
i\partial_+\Sigma)]\Sigma^{-1}\nonumber\\
&-&\frac{1+C_V}{4\pi}\Sigma i\partial_-
\Sigma^{-1}
+\frac{1}{4\pi}Ui\partial_-U^{-1}+\{b_-^{(0)},c_-^{(0)}\}.
\end{eqnarray}
These conditions must be implemented on the Hilbert space, and
require the physical Hilbert space to be neutral with respect to the BRST
charges.\footnote{ Within the context of the generating functional this is insured
by coupling {\it ab initio}
the sources to the set of intrinsic fields
defining the model, as done in (\ref{generatingZ}). In this way we ensure to reproduce all
Wightman functions defining the model.}  In other words: the field $U$ is
not a BRST invariant. We can nevertheless arrive at a factorization
in terms of BRST invariant sectors as follows.

Since $U$ and $G$ are WZW fields describing non-interacting fermions,
we may factorize them into right and left moving parts,
$U=U_rU_{\ell}, G=G_rG_{\ell}$. The Poloyakov-Wiegmann formula \cite{PW}
\be\label{PW}
\Gamma[AB]=\Gamma[A] + \Gamma[B] +
\frac{1}{4\pi} \int d^2x tr(A^{-1}\partial_+A)(B\partial_-B^{-1})
\ee
allows us to write (recall (\ref{factorizedZhat}))
\bear
{\cal Z}^{^{(0)}}_{_F}\,\int \,{\cal D} U\,e^{\,i \Gamma [U]} &&=
{\cal Z}^{^{(0)}}_{_{U(1)}}\,\int \,{\cal D} G\,
e^{\,i \Gamma [G]}\,\int \,{\cal D} U\,e^{\,i \Gamma [U]}\nonumber\\
&&= {\cal Z}^{^{(0)}}_{_{U(1)}}\,\int \,{\cal D} \widetilde{U}\,
e^{\,i \Gamma [\widetilde{U}]}\,\int
\,{\cal D}
\widetilde{G}\,e^{\,i \Gamma [\widetilde{G}]} =\nonumber\\
&& = {\widetilde{\cal Z}}^{^{(0)}}_{_F}\, \int \,{\cal D} \widetilde{U}\,
e^{\,i \Gamma [\widetilde{U}]}
\ear
where ${\cal Z}^{^{(0)}}_{_{U(1)}}$ carries the $U(1)$ degrees of
freedom of the original free
fermions $\psi^o$, $ \widetilde{U} \doteq G_r U_\ell $\,
$ \widetilde{G} \doteq U_r G_\ell $ , and where we have taken account of
the right-(left-) moving property of $G_r,U_r$ ($G_\ell,U_\ell$) in the
decomposition $G=G_rG_{\ell}$, $U=U_rU_{\ell}$. We see
from (\ref{constr+}) and (\ref{constr-}) that the new field
$\widetilde U$ is BRST neutral, $ \Big [ {\cal Q}_{_{BRST}},\widetilde{U}
\Big ] = 0 $.

The partition function (\ref{factorizedZhat}) can thus be
factorized into two partition functions that are separately neutral with
respect to the BRST charges:
\be\label{Zfact}
Z\,=\,\Big ( \int {\cal D} \widetilde{U}\,e^{\,i \Gamma [\widetilde{
U}]} \Big )\,\Big (\,
{\cal Z}_{_{gh}}^{^{(0)}}\,\widetilde {\cal Z}^{^{(0)}}_{_F}\,
{\cal Z}^{^\Sigma} \Big )_{QCD_2}\,.
\ee

The above factorization of the partition function
suggests that the Hilbert space of the anomalous chiral theory can be
factorized
as $ \mbox{\boldmath ${\cal H}$} =
\mbox{\boldmath ${\cal H}$}_{\widetilde U} \otimes
\mbox{\boldmath ${\cal H}$}_{_{QCD_2}} $, implying
that $ \mbox{\boldmath ${\cal H}$} $ contains
$ \mbox{\boldmath ${\cal H}$}_{_{QCD_2}} $ as a {\it physical}
subspace. However, as we now show, this still is
an improper factorization of the Hilbert space since violates certain
superselection rules. This is contained in the following proposition
\cite{MPS,Bel}:

{\it Let ${\cal Q}$ a local charge operator satisfying
\be
\big [ {\cal Q},\mbox{\boldmath $\Im$} \big ] = 0\,,
\ee
and which is trivialized in the restriction
from $\mbox{\boldmath ${\cal H}$}^{^B}$ to
$\mbox{\boldmath ${\cal H}$}\,\subset\, \mbox{\boldmath ${\cal H}$}^{^B}$, i. e.,
\be
{\cal Q}\, \mbox{\boldmath ${\cal H}$}^{^B} \neq 0 \,\,,\,\, {\cal Q}\,
 \mbox{\boldmath ${\cal H}$} = 0\,.
\ee

Let $\mbox{\boldmath ${\cal A}$}$ be an operator
satisfying $ \big [ {\cal Q}_{_{BRST}},$\mbox{\boldmath ${\cal A}$}$ \big ] = 0 $, but carrying
the charge ${\cal Q}$ such that $\big [ {\cal
Q},$\mbox{\boldmath ${\cal A}$}$ \big ] \neq 0 $. Then the
operator $\mbox{\boldmath ${\cal A}$}$  does not belong to the
intrinsic local field algebra $\mbox{\boldmath $\Im$}$ and cannot be defined
as a solution of the BRST condition in
$\mbox{\boldmath ${\cal H}$}$, i. e., $\mbox{\boldmath ${\cal A}$}$ is not an
element of $\mbox{\boldmath ${\cal H}$}$.}

We now show that the BRST invariant field $\widetilde U$ carries a charge that
is trivialized in the restriction
from $ \mbox{\boldmath ${\cal H}$}^{^B}$ to $\mbox{\boldmath ${\cal H}$}$. To
this end consider the left and right WZW currents
\bear
 j^\mu_\ell &=& - \frac{1}{4\pi}
 U^{-1} i( \partial ^\mu + \tilde \partial ^\mu ) U\,,\nonumber\\
j^\mu_r &=& -\frac{1}{4\pi}
  U i( \partial ^\mu - \tilde \partial ^\mu ) U^{-1}\,.
\ear
The conserved vector and axial vector currents are respectively defined by
$$ j^\mu = \frac{1}{2} ( j^\mu_r + j^\mu_\ell )\,, $$
\be
\widetilde{j}^\mu = \frac{1}{2} ( j^\mu_r - j^\mu_\ell ) = \epsilon^{\mu \nu}
j_\nu\,.
\ee

Denoting by ${\cal Q}$ and ${\cal Q}^{^5}$ the respective charges, we have
\be
\big [ {\cal Q}^{^5},\mbox{\boldmath $\Im$} \big ] =  0\,\,,\,\,
{\cal Q}^{^5}\,\mbox{\boldmath ${\cal H}$} = 0\,.
\ee
since $[ {\cal Q}^{^5},U \big ]= 0\,$.
The field $U$ is however not BRST invariant, and therefore does belong to the
field algebra $\mbox{\boldmath $\Im$}$. This shows that \mbox{\boldmath ${\cal H}$} cannot be factorized as $ \mbox{\boldmath ${\cal H}$} =
\mbox{\boldmath ${\cal H}$}_{U} \otimes
\mbox{\boldmath ${\cal H}$}_{_{QCD_2}} $.
On the other hand, $\widetilde U$ is BRST invariant, but
\be
\big [ {\cal Q}^{^5}, \widetilde U \big ] = \widetilde U\,.
\ee
The BRST invariant field $\widetilde U$ thus carries the charge
${\cal Q}^{^5}$, and therefore again does
not belong to the field algebra $\mbox{\boldmath $\Im$}$ according to the above proposition. We conclude that the physical
Hilbert space cannot be factorized in the form $\mbox{\boldmath ${\cal H}$}_{\widetilde U} \otimes
\mbox{\boldmath ${\cal H}$}_{_{QCD_2}} $, either; i.e., just as in the abelian
case \cite{Bel}, the chiral $QCD_2$ for the JR
parameter $a=2$ is not equivalent to $QCD_2$ plus a ``decoupled''
free massless field. In Appendix A we illustrate by
an explicit example the effect of the use of an
external field algebra in deriving basic physical properties of the
model and show that the improper factorization of the
Hilbert space leads to misleading conclusions about the physical
content of the anomalous chiral theory.

\section{Extended Gauge Invariant Local Formulation}
\setcounter{equation}{0}

Our starting point is the partition function (\ref{ZGI}) in the local
$GNI$ formulation. The action $S[U,V]$ is not invariant under the gauge
transformation $U\to Ug,V\to g^{-1}V$:
\be\label{SGI1}
S[U,V]\to S[Ug,g^{-1}V]
=S[U,V]+S_{WZ}[U,V,g]\,,\ee
with $S_{WZ}[U,V,g]$ the Wess-Zumino-Witten action \cite{WZW}
\bear\label{1.4}
S_{WZ}[U,V,g]&=&\frac{a}{2}\Gamma[g]+\left(\frac{a}{2}-1\right)
\Gamma[g^{-1}]\nonumber\\
&&+\frac{a}{2}\frac{1}{4\pi}\int d^2x\ { {tr}}\left (U^{-1}i\partial_+Ugi
\partial_-g^{-1}\right )\nonumber\\
&&+\left (\frac{a}{2}-1\right )\frac{1}{4\pi}\int d^2x\ { {tr}}\left (
g i \partial_+g^{-1} V_i\partial_-V^{-1}\right )\,.\ear

We proceed now to embed the anomalous theory into a gauge
theory, following the
standard procedure in configuration space \cite{HaTsu}. To this end
we introduce in the generating functional $ {\cal Z} [\bar
\eta,\eta,J_\mu] $, given by (\ref{generatingZ}), the identity
\be
\Delta_{\cal F}[UV]\int {\cal D}g\delta[{\cal F}(Ug^{-1},gV)]=1\,.
\ee
The Faddeev-Popov determinant $\Delta_{\cal F}[UV]$ is
gauge-invariant. Setting $Ug^{-1}=\overline U,\ gV=\overline V$,
using the invariance of the Haar measure, we obtain from (\ref{generatingZ}) 
the generating functional of the $GI$ formulation

$$ {\cal Z}_{_{GI}} [\bar \eta,\eta,J_\mu] = {\cal Z}^o_{_F}
\int {\cal D}g \int {\cal D} \overline U {\cal D}\overline V \Delta_{\cal F}
[\overline U\overline V] \delta[{\cal F}(\overline U,\overline V)] \,
e^{iS[\overline Ug,g^{-1}\overline V]} $$
\be\label{ZGI}
\times e^{\, i \int d ^2 z \Big \{
\eta^\dagger_\ell g^{-1} \overline V \psi^o_\ell + \Psi^{o \dagger}_\ell
{\overline V}^{-1} g\, \eta_\ell +  \frac{i}{e} J_-
\Big ( (\overline U g )^{-1} \partial_+ (\overline U g)
\Big ) + \frac{i}{e} J_+ \Big ( (g^{-1} \overline V) \partial_- (g^{-1}
\overline V)^{-1} \Big ) \Big \}}\,.
\ee
Repeating this procedure one can easily generalize this expression to
more general gauges. Dropping ``bars'', etc., the generalization reads
\be
{\cal Z}_{_{GI}} [\bar \eta,\eta,J_\mu] = {\cal Z}^o_{_F}\,\int {\cal D} g \int
{\cal D} U{\cal D} V \Delta_{\cal F}[Ug,g^{-1}V] \delta[{\cal F}(U,V,g)]
\times e^{\,iS[Ug,g^{-1}V]} \times
\ee
\be e^{\,i \int d ^2 z \Big \{
\eta^\dagger_\ell g^{-1} V \psi^o_\ell + \psi^{o \dagger}_\ell
{V}^{-1} g \eta_\ell + \frac{i}{e} J_-
\Big ( (U g )^{-1} \partial_+ (U g)
\Big ) + \frac{i}{e} J_+ \Big ( (g^{-1}) \partial_- (g^{-1} V)^{-1} \Big ) \Big \}}\,,
\ee
where
\be
S[Ug,g^{-1}V] =  S_{_{_{YM}}}[UV]\,-\,\Big ( C_{_V} + \frac{a}{2} \Big)\,
\Gamma[UV] \, + \frac{a}{2} \Gamma [Ug] + \Big (\frac{a}{2} - 1
\Big ) \Gamma [g^{-1}V]\,,
\ee
and which also can be written in terms of the WZ action as (\ref{SGI1}).

In the unitary gauge, ${\cal F} (U,V,g) = g - 1$, and we recover the
$GNI$ formulation. However, as exhaustively discussed
in Refs. \cite{Bel,GR,AAR}, the isomorphism between these two
formulations is valid in an arbitrary gauge.

\subsection{CONSTRAINT STRUCTURE}

In the $GNI$ bosonized decoupled formulation, the
action $S[U,V]$ describes from the Dirac point of view an unconstrained
system. The second-class constraints of the original quantum
fermionic formulation have now been replaced by BRST constraints
restricting the bosonic
Hilbert space of the present formulation to the physical Hilbert
space $\mbox{\boldmath ${\cal H}$}$ generated by the intrinsic set of
field operators $\{\psi^o_r,\psi_\ell,{\cal A}_\mu\}$ which defines
the original theory. On the other hand, in
the $GI$ bosonized formulation, we expect $S[Ug,g^{-1}V]$ to describe a
Hamiltonian system with only one
first-class constraint generating the extended gauge transformation $U\to UG,\ V\to
G^{-1}V,g\to G^{-1}g$. In the following we demonstrate this. Because of the
non-abelian nature of the problem, the demonstration
involves some technicalities.

Our starting point is the $GI$ effective bosonized action
\be\label{GIA}
S[Ug,g^{-1}V]=S_{YM}[\Sigma] - \Big (\frac{a}{2} + C_V \Big ) \Gamma[\Sigma] +
\frac{a}{2} \Gamma[Ug] + \Big ( \frac{a}{2} - 1 \Big ) \Gamma [
g^{-1} V]\,, \ee
obtained by performing a gauge transformation in the bosonized action (\ref{GNIA}).

The general form of the constraint can be displayed as follows: use
that $ V = U^{-1} \Sigma$, and express the action (\ref{GIA}) in
terms of the gauge invariant field variables $\Sigma$ and $\overline U = U g $: $ S
[Ug,g^{-1}V] \equiv S [\Sigma,\overline U]$. Thus we have for the
canonical momenta
\be\label{canmomenta}
\Pi^{(U)}_{ij} = \Pi^{(\overline U)}_{il}\, g_{jl}\,,\quad
\Pi^{(g)}_{ij} = \Pi^{(\overline U)}_{lj} U_{li}\,,
\ee
with
\be
\Pi^{(\overline U)}_{il} \doteq \frac{\delta S [\Sigma,\overline U]}
{\delta (\partial_0 \overline U)_{il}}\,.
\ee
Note that the product of the Bose
field variables $U$, $V$, and the corresponding ``transposed'' momenta is
a gauge invariant quantity
\be\label{giconstraint}
\widetilde\Pi^{(U)}U = \widetilde\Pi^{(\overline U)}
\overline U \,,
\ee
where the ``tilde'' stands for ``transpose''. From (\ref{canmomenta})
and (\ref{giconstraint}) we read off the
general form of the primary constraint
\be\label{1.13}
\Omega :=\widetilde\Pi^{(U)}U-g\widetilde\Pi^{(g)}=0\,,
\ee

The general conclusions of this section are valid for arbitrary JR parameter
$a$. In order to simplify the discussion we choose $a=2$. In this case
(\ref{GIA}) reduces to
\be\label{SGI}
S[Ug,g^{-1}V]=S_{YM}[\Sigma]-(1+C_V)\Gamma[\Sigma]+
\Gamma[Ug]\,.
\ee
The canonical quantization can proceed in
two ways:

i) We make use of the Polyakov-Wiegmann formula \cite{PW}
\be\label{PW1}
\Gamma[Ug]=\Gamma[U]+\Gamma[g]+
\frac{1}{4\pi}\int d^2x\ {tr} [ (U^{-1}\partial_+U)(g\partial_-g^{-1})]\,,
\ee
and note that $\Gamma[G]$ is of the form \cite{AR}
\be\label{Gamma}
\Gamma[G]=\frac{1}{8\pi}\int d^2x\ {tr} ( \partial_\mu
G\partial^\mu G^{-1} ) + \frac{1}{4\pi} \int d^2x\ {tr}(A(G)\partial_0G)\,.
\ee
We then obtain for the momenta canonically conjugate to $U$ and $g$
\bear\label{pi}
\pi^{(U)}_{ij}=\frac{1}{4\pi}\left\{\partial_0U^{-1} +
A (U)+[(g\partial_-g^{-1})U^{-1}] \right\}_{ji}\,,\nonumber\\
\pi^{(g)}_{ij}=\frac{1}{4\pi}\left\{\partial_0g^{-1} + A (g) +
[g^{-1}(U^{-1}\partial_+U)] \right\}_{ji}\,.
\ear

ii) We leave (\ref{SGI}) as it stands and compute the
momenta conjugate to $U$ and $g$. Making use of (\ref{Gamma}) we then
obtain
\bear\label{Pi}
\Pi^{(U)}_{ij}=\frac{1}{4\pi}\left\{\partial_0U^{-1} +
[(g\partial_0g^{-1})U^{-1}+gA(Ug)] \right\}_{ji}\,,\nonumber\\
\Pi^{(g)}_{ij}=\frac{1}{4\pi}\left\{\partial_0g^{-1}-
[g^{-1}(U^{-1}\partial_0U)+A(Ug)U] \right\}_{ji}\,..\ear

Making use of fundamental property \cite{AR}
\be\label{A}
\frac{\partial A(G)_{ij}}{\partial G_{kl}}-
\frac{\partial A(G)_{lk}}{\partial G_{ji}}=\partial_0G^{-1}_{ik}
G^{-1}_{lj}-G^{-1}_{ik}\partial_0 G^{-1}_{lj}\,,
\ee
it is straightforward to show that the two sets of canonical
momenta are related by a canonical transformation. We leave
the demonstration of this to the appendix B.

In the following it turns out to be more convenient to work
with the canonical momenta (\ref{Pi}). Defining $\Omega^a={tr}
\,(t^a\Omega)$, with $[t^a,t^b]=if^{abc}\,t^c$, a simple calculation
shows that
\be\label{1.16}
\left\{\Omega^a(x),\Omega^b(y)\right\}=-f^{abc}\,\Omega^c(x)\delta(x^1-y^1).\ee
Hence the primary constraints (\ref{1.13}) are first class \cite{Di}.
It remains to show that there are no further constraints. To see this
we need to compute $\{\Omega^a,H_T\}$, where $H_T$ is the total Hamiltonian,
$H_T=H_c+\int d^2xv^a\Omega^a$. It is a straightforward matter to
show that $H_c$ is weakly equivalent to
\bear\label{1.17}
H_c\approx\int dy^1 {tr}\Bigl\{&&-\widetilde\Pi_{(E)}\partial_1 E
+{\widetilde{\hat\Pi}}_{(\Sigma)}\partial_1\Sigma
+2ie\widetilde{\hat\Pi}_{(\Sigma)}\widetilde\Pi_{(E)}+\frac{1}{2}E^2\nonumber\\
&&+\frac{(1+C_V)}{2\pi}e^2(\widetilde\Pi_{(E)})^2
-4\pi(\widetilde{\hat\Pi}_{(U)} U)(g\widetilde{\hat\Pi}_{(g)})\nonumber\\
&&-\frac{1+C_V}{4\pi}\partial_1\Sigma\partial_1\Sigma^{-1}
+\frac{1}{8\pi}\partial_1(Ug)\partial_1(Ug)^{-1}\Bigr\}\,,\ear
where (see also \cite{AR1})
\bear\label{1.18}
&&\widetilde{\hat\Pi}_{(U)}=\widetilde \Pi_{(U)}-\frac{1}{4\pi}gA(Ug)\,,
\nonumber\\
&&\widetilde{\hat\Pi}_{(g)}=\widetilde \Pi_{(g)}-\frac{1}{4\pi}A(Ug)U\,.\ear

For the computation of $\{\Omega^a,H_c\}$ it is useful
to observe that $\int d^2x\epsilon^a\Omega^a$ is the generator of
the gauge transformation $U\to UG,\ V\to G^{-1}V$,
$g\to G^{-1}g$:
\bear\label{1.19}
&&\{U_{ij}(x),\Omega^a(y)\}=(Ut^a)_{ij}\delta(x^1-y^1)\,,\nonumber\\
&&\{g_{ij}(x),\Omega^a(y)\}=-(t^ag)_{ij}\delta(x^1-y^1)\,.\ear
As a consequence we have for any functional of $Ug$,
\be\label{f[Ug]}
\{f[Ug],\Omega^a(x)\}=0\,.
\ee
Furthermore
\bear\label{1.21}
&&\{\widetilde\Pi_{ij}^{(U)}(x),\Omega^a(y)\}=-(t^a\widetilde\Pi^{(U)})_{ij}
\delta(x^1-y^1)\,,\nonumber\\
&&\{\widetilde\Pi_{ij}^{(g)}(x),\Omega^a(y)\}=(\widetilde\Pi^{(g)} t^a)_{ij}
\delta(x^1-y^1)\,.\ear
>From here it follows that
\be\label{1.22}
\left\{\{\Omega^a,{tr}\left(\widetilde{\hat\Pi}_{(U)}
Ug\widetilde{\hat\Pi}_{(g)}\right)\right\}=0.\ee
This result, together with (\ref{f[Ug]}) then implies $\{\Omega^a,
H_c\}=0$, which shows that $\Omega^a=0$ are the only constraints.

\section{Conclusions and Final Remarks}

The bosonization of chiral $QCD_2$ in terms of group valued fields involves
a Hilbert space ${\bf{\cal H}}^B$ which is much larger than the
physical Hilbert space ${\bf{\cal H}}$. This required a careful analysis
based on the construction of the physical Hilbert space as a representation of the intrinsic field algebra, in order to avoid misleading conclusions. We have theirby shown that the BRST
conditions to be imposed on ${\bf{\cal H}}^B$ are in general not sufficient
to define ${\bf{\cal H}}$, and that certain superselection rules also have
to be taken into account. This observation proved crucial for establishing
the non-equivalence of chiral $QCD_2$ with JR-parameter $a=2$ to $QCD_2$.

The above analysis was carried out in the GNI formulation,
where the bosonic degrees of freedom are not restricted by Dirac-type constraints \cite{Di}. In the GNI formulation, we have an unconstrained system in the sense of Dirac and the BRST conditions replace
the second class
constraints of the original formulation in terms of
gauge and fermion fields.
When embedding the theory into a gauge theory
by suitable addition of a WZW-term, following the procedure of
\cite{HaTsu}, we showed that one arrives at a constrained system in the sense
of Dirac, with one set of first class constraints. This was to be
expected, since the embedding has left us with a formulation exhibiting
a local symmetry. As we showed, these first class constraints are indeed
the generators of gauge transformations in the bosonic formulation.

An open question refers to the
generalization of the bosonization technique for obtaining also a decoupled
formulation of chiral $QCD_2$ in the  case $a>1$. It appears
questionable at present, whether a  complete factorization of the
partition function can also be achieved in the general case.

\newpage

\section*{Appendix A}
\renewcommand{\theequation}{A.\arabic{equation}}
\setcounter{equation}{0}
In this Appendix we examine the effect of the use of a redundant
field algebra in deriving basic physical properties of the model. In analogy with the abelian case \cite{CW,Bel}, this will be
illustrated by an example in which the Bose field $\widetilde U$ is
improperly introduced in the physical field algebra.

To begin with, consider the chiral density operator (normal ordering and point-splitting are presumed)
\be
{\cal M} \doteq {tr} (\psi_\ell\, \psi^{o \dagger}_r ) \,\in\,
\mbox{\boldmath $\Im$} \,,
\ee
\be
\Big [ {\cal Q}_{_{BRST}},{\cal M} \Big ] = 0\,.
\ee
In terms of the bosonic variables we can write
\be
{\cal M} = {tr} (V \psi_\ell^o \psi^{o \dagger}_r) = {tr} (V G^{-1}) = {tr}
( U^{-1} \Sigma G^{-1} )\,.
\ee
where $G$ is a WZW field.

In order to decouple the chiral operator into two BRST-neutral
pieces, consider the operator
\be
{\cal M} = (G_r U_\ell)^{-1} G_r U_r^{-1} \Sigma
G^{-1} = \widetilde{U}^{-1}\,( G_r U_r^{-1} \Sigma G^{-1})
= \widetilde U^{-1}(\widetilde G^{-1}\Sigma)\,.
\ee
where $ \widetilde U = G_r U_\ell $. Defining a ``new'' chiral density operator by {\it extracting} the
$\widetilde{U}$ dependence in (A.4),
\be
\widetilde{\cal M} \doteq {tr} (\Sigma \widetilde{G}^{-1})\,,
\ee
with $ \widetilde G = U_r G_\ell $. The chiral density (A.5)
corresponds to the mass operator of $QCD_2$: $ \widetilde{\cal M}\,
=\, {tr} (\Sigma \widetilde{G}^{-1})$.

The definition of the
operator $\widetilde {\cal M}$ relies on the incorrect assumption that the
field $\widetilde U$ can be introduced in the intrinsic field algebra
$\mbox{\boldmath$\Im$}$. Since the operator $\widetilde U$ carries the
charge $ {\cal Q}^{^5} $, that is trivialized in the restriction
from $\mbox{\boldmath ${\cal H}$}^{^B}$ to $\mbox{\boldmath ${\cal H}$}$, one
cannot define the operator $\widetilde U$ as a solution of the
subsidiary condition on $\mbox{\boldmath${\cal H}$}$ and therefore
the operator $\widetilde {\cal M}$ also cannot be
defined on $\mbox{\boldmath${\cal H}$}$.

\newpage
\section*{Appendix B}
\renewcommand{\theequation}{B.\arabic{equation}}
\setcounter{equation}{0}
We show here that the two definitions (\ref{pi}) and (\ref{Pi}) of the
momenta conjugate to $U$ and $g$ are canonically equivalent
by showing that the canonical commuation relations of
$\Pi^{(U)}_{ij},U_{ij},\Pi^{(g)}_{ij},g_{ij}$
 imply canonical
commutation relations of $\pi^{(U)}_{ij},U_{ij},\pi^{(g)}_{ij},
g_{ij}$. From (\ref{pi}) and (\ref{Pi}) we have
\bear\label{B.1}
&&\tilde\pi^{(U)}=\tilde\Pi^{(U)}+\frac{1}{4\pi}
(A(U)-gA(Ug))-\frac{1}{4\pi}(g\partial_1g^{-1})U^{-1}\,,\\
&&\tilde\pi^{(g)}=\tilde\Pi^{(g)}+\frac{1}{4\pi}
(A(g)-A(Ug)U)-\frac{1}{4\pi}g^{-1}(U^{-1}\partial_1U).\nonumber\ear
It is clear that (\ref{B.1}) implies
\bear\label{B.2}
\{U_{ij}(x)\Pi^{(U)}_{kl}(y)\}&=&\{U_{ij}(x)\Pi^{(U)}_{kl}(y)\}
=\delta_{ik}\delta_{jl}\delta(x^1-y^1)\,,\nonumber\\
\{g_{ij}(x),\Pi_{kl}^{(g)}(y)\}&=&\{g_{ij}(x),\Pi_{kl}^{(g)}(y)\}
=\delta_{ik}\delta_{jl}\delta(x^1-y^1)\,,\nonumber\\
\{U_{ij}(x),\Pi_{kl}^{(g)}(y)\}&=&\{g_{ij}(x),\Pi^{(U)}_{kl}
(y)\}=0\,.\ear
We now compute the remaining commutators.\\

i) For $\{\pi_{ij}^{(U)},\pi_{kl}^{(U)}\}$ one finds,
using (\ref{B.1}),
\bear\label{B.3}
4\pi\{\pi_{ij}^{(U)}(x),\pi_{kl}^{(U)}(y)\}&=&
\Bigl\{-\left(\frac{\partial A_{lk}(U)}
{\partial U_{ij}}-\frac{\partial A_{ji}(U)}
{\partial U_{kl}}\right)\nonumber\\
&+&\left(g_{lr}\frac{\partial A(Ug)_{rk}}{\partial U_{ij}}-
g_{jr}\frac{\partial A(Ug)_{ri}}{\partial U_{kl}}\right)\nonumber\\
&-&\left[\left(g\partial_1g^{-1}\right)U^{-1}\right]_{li}U^{-1}_{jk}\nonumber\\
&+&\left[\left(g\partial_1g^{-1}\right)U^{-1}\right]_{jk}U^{-1}_{li}\Bigr\}
\delta(x^1-y^1).\ear
Noting that ($\bar U = Ug$)
\be\label{B.4}
g_{lr}\frac{\partial A(\bar U)_{rk}}
{\partial U_{ij}}=g_{lr}\frac{\partial A(\bar U)_{rk}}
{\partial\bar U_{is}}g_{js}\,,\ee
we find
\begin{eqnarray}\label{B.5}
g_{lr} \frac{\partial A(\bar U)_{rk}}{\partial U_{ij}} -
g_{jr} \frac{\partial A(\bar U)_{ri}}{\partial U_{kl}} &&=
[\frac{\partial A(U)_{lk}}{\partial U_{ij}} -
\frac{\partial A(U)_{ji}}{\partial U_{kl}}]\nonumber\\
 &&+[(g\partial_1 g^{-1})U^{-1}]_{li} ]U_{jk} -
U^{-1}_{li}[ (g\partial_1 g^{-1}) U^{-1}]_{jk}.
\end{eqnarray}
Making further use of (\ref{B.3}), one finds that the r.h.s.
of (\ref{B.3}) vanishes identically. Similarly one verifies
that
$\{\pi_{ij}^{(g)}(x),\pi^{(g)}_{kl}(y)\}=0$.\\

ii) The computation of $\{\pi_{ij}^{(U)}(x),\pi^{(g)}_{kl}(y)\}=0$
follows along similar lines. We find
\bear\label{B.6}
4\pi\{\pi_{ij}^{(U)}(x),\pi^{(g)}_{kl}(y)\}
&&=\Bigl\{\left(\frac{\partial A(\bar U)_{lr}}{\partial U_{ij}}U_{rk}
-g_{jr}\frac{\partial A(\bar U)}{\partial g_{kl}}\right)\nonumber\\
&&-(g^{-1}U^{-1})_{li}(U^{-1}\partial_1U)_{jk}
+(g\partial_1g^{-1})_{jk}(g^{-1}U^{-1})_{li}\Bigr\}\nonumber\\
&&-(g^{-1}(y)U^{-1}(y))_{ri}\delta_{jk}\partial^x_1\delta(x^1-y^1)\nonumber\\
&&+(g^{-1}(x)U^{-1}(x))_{li}\delta_{jk}\partial^x_1\delta(x^1-y^1)\,.
\ear
Note that the sum of the last two terms in (\ref{B.6}) can be written as
$-\delta_{jk}\partial_1(Ug)^{-1}\cdot \delta(x^1-y^1)$. Noting futher that $(\bar U=Ug)$
\be\label{B.7}
\frac{\partial A(\bar U)_{er}}{\partial U_{ij}}U_{rk}-g_{jr}
\frac{\partial A(\bar U)_{ri}}{\partial g_{kl}}=
g_{js}\left(\frac{\partial A(\bar U)_{lr}}
{\partial \bar U_{is}}-\frac{\partial A(\bar U)_{si}}
{\partial \bar U_{rl}}\right)U_{rk}\,,\ee
and using (\ref{A}), we find that the r.h.s. of (\ref{B.5})
vanishes identically. This concludes the proof.

\underline{{\it Acknowledgments}}: The authors are grateful to the
Conselho Nacional de Desenvolvimento
Cient\'{\i}fico e Tecnol\'ogico (CNPq - Brasil) and DAAD
for partial financial support.

\end{document}